\documentclass[copyright,creativecommons,noderivs]{eptcs}

\providecommand{\event}{GandALF 2011} 
\usepackage{breakurl}               

\usepackage[utf8]{inputenc}
\usepackage[T1]{fontenc}
\usepackage[english]{babel}

\usepackage{relsize}

\usepackage{amsmath,amssymb}
\usepackage{amsthm}
\usepackage{mathtools}

\usepackage{booktabs}
\usepackage{multirow}

\usepackage{tikz}
\tikzstyle{strong node}=[ultra thick]
\tikzstyle{deleted node}=[densely dashed]
\tikzstyle{deleted edge}=[densely dashed]
\tikzstyle{replacement}=[draw,rectangle,ultra thick,minimum size=0.5cm]
\tikzstyle{dynnets style}=[node distance=1.5cm, auto, semithick]

\usepackage[all]{hypcap}

\newcommand{\NP}{\textsc{NP}}
\newcommand{\PSPACE}{\textsc{PSpace}}
\newcommand{\EXPTIME}{\textsc{ExpTime}}
\newcommand{\TwoEXPTIME}{\textsc{2ExpTime}}
\newcommand{\cG}{\ensuremath{\mathcal{G}}}
\newcommand{\dotcup}{\mathbin{\mathaccent\cdot\cup}}

\newcommand{\lr}{\langle\,}
\newcommand{\rr}{\,\rangle}
\newcommand{\blank}{{\mathsmaller{\llcorner}\!\mathsmaller{\lrcorner}}} 
\newcommand{\relabel}{\longrightarrow}
\newcommand{\move}{\xrightarrow{\text{move}}}
\newcommand{\create}[1][]{\xrightarrow{\ensuremath{\text{create}\mathsmaller{#1}}}}
\newcommand{\createstrong}[1][]{\xrightarrow{\ensuremath{\text{s-create}\mathsmaller{#1}}}}

\newcommand{\Theorem}{Theorem~}
\newcommand{\Lemma}{Lemma~}

\newcommand{\Remark}{Remark~}

\newcommand{\Sect}{Section~} 
\newcommand{\Fig}{Figure~} 
\newcommand{\Table}{Table~}

\newtheorem{theorem}{Theorem}[section]

\newtheorem{lemma}[theorem]{Lemma}

\theoremstyle{definition}

\theoremstyle{remark}
\newtheorem{remark}[theorem]{Remark}
\newtheorem{example}[theorem]{Example}

\title{Connectivity Games over Dynamic Networks}

\author{
Sten Gr\"{u}ner\thanks{This author was supported by the DFG Research Training Group ``AlgoSyn'' (Algorithmic Synthesis of Reactive and Discrete-Continuous Systems), German Research Foundation grant DFG~GRK~1298.}
\institute{Chair of Process Control Engineering\\
RWTH Aachen University}
\email{s.gruener@plt.rwth-aachen.de}
\and
Frank G.\ Radmacher\thanks{This author was supported by the DFG Cluster of Excellence ``UMIC'' (Ultra High-Speed Mobile Information and Communication), German Research Foundation grant DFG~EXC~89.} \qquad Wolfgang Thomas
\institute{Chair of Computer Science 7\\
RWTH Aachen University}
\email{\{radmacher,thomas\}@automata.rwth-aachen.de}
}
\def\titlerunning{Connectivity Games over Dynamic Networks}
\def\authorrunning{S.~Gr\"{u}ner, F.G.~Radmacher \& W.~Thomas}

\begin{document}

\maketitle

\begin{abstract}
  A game-theoretic model for the study of dynamic networks is analyzed.
  The model is motivated by communication networks that are subject to failure of nodes and where the restoration needs resources.
  The corresponding two-player game is played between ``Destructor'' (who can delete nodes) and ``Constructor'' (who can restore or even create nodes under certain conditions).
  We also include the feature of information flow by allowing Constructor to change labels of adjacent nodes.
  As objective for Constructor the network property to be connected is considered, either as a safety condition or as a reachability condition (in the latter case starting from a non-connected network).
  We show under which conditions the solvability of the corresponding games for Constructor is decidable, and in this case obtain upper and lower complexity bounds, as well as algorithms derived from winning strategies.
  Due to the asymmetry between the players, safety and reachability objectives are not dual to each other and are treated separately.
  \par\addvspace\baselineskip
\end{abstract}

\section{Introduction and Motivation}

A classical scenario for the application of game-theoretic methods in verification is the antagonism between a possibly malicious \emph{environment} and a \emph{system} (or its control component) that has to guarantee a desired behavior given any choice of actions of the \emph{environment}.
The task of verification is then to show that in this game between \emph{environment} and \emph{system} the player \emph{system} has a winning strategy, and in an ideal situation it should even be possible to generate such a winning strategy from the specification of the desired behavior (defined, e.g., in terms of a formula of temporal logic).

The present paper pursues this view in a specific context that is of central interest in the theory of communication networks.
We study the antagonism between ``suppliers'' and ``users'' of a communication network on one side and the generation of faults (either by nature or by malicious interference) on the other.
So, we consider \emph{dynamic network games} in which a game position is just given by a current shape of a network.
The party that generates faults is modeled by a player called \emph{Destructor} who can ``delete'' nodes in a network.
In the present paper we only consider changes in the set of nodes (and induced changes in the set of edges).
The more general case that arises when including edges in the dynamics involves heavier notation, but does not affect the general results as they appear in this paper.
(More precisely, our framework is able to simulate edge deletions by modeling each edge as a vertex~\cite{Gruener2011-diploma}.)

The other parties involved are the suppliers of the network and the users.
There are many ways to model these parties.
We consider here a model that represents a compromise between conceptual simplicity and adequacy for practical applications%
\footnote{We thank our colleagues in the research cluster UMIC (Ultra High-Speed Mobile Information and Communication) for their contributions in devising the current model.}.
The aspect of simplicity is introduced by a merge of the two parties suppliers and users in a single player called \emph{Constructor}.
This player has the power to restore deleted nodes or even to create new nodes (i.e., to extend the network beyond its original shape), matching the purpose of a supplier, and she also can transmit information along edges of the network, matching the actions of an user.
The latter feature is realized by the possibility to relabel two nodes that are connected by an edge.

A detailed description of these possible actions by Destructor and Constructor yields a dynamic network game in which these players carry out their moves in alternation, step by step changing the shape and the labeling of the network.
In the present paper we analyze these games only with one objective (winning condition for Constructor), namely with the objective to guarantee that the network is connected. The objective arises in two versions:
as a safety game in which connectivity of the network is to be guaranteed forever by Constructor, or as a reachability game in which Constructor has to construct a connected network, starting from a disconnected one.
Since we assume complete information, these games are trivially determined.
Our aim is to clarify under which assumptions these games are effectively solvable, i.e., that one can decide who wins (and in this case to construct a winning strategy for the winner).
One should note that by the independent and very different conceptions of the two players, there is no direct duality between reachability and safety objectives; both games have to be analyzed separately.
On the other hand, a simplification is built into our model by our decision only to declare connectivity of the network as Constructor's objective, i.e., that the aim of the users to realize the transfer of information from certain source nodes to target nodes is not taken into account.
(For a study of network games including this aspect, see the paper~\cite{GroRadTho10}.)%
\begin{figure}%
  \centering
  \parbox[b]{0.45\textwidth}{%
  \centering
  \begin{tikzpicture}[style=dynnets style]
    \tikzstyle{every node}=[minimum size=0.7cm]
    
    \node[circle,draw,style=deleted node]   (v1) at (-0.75,0)   {$v$};
    \node[circle,draw,style=strong node]    (u1) [left of=v1,node distance=1.0cm]   {$u$};
    
    \draw[->] (0,0) -- (0.5,0);
    
    \node[circle,draw]                      (u2) at (1.25,0)    {$u$};
    \node[circle,draw,style=strong node]    (v2) [right of=u2,node distance=1.0cm]  {$v$};
    
    \path[style=deleted edge] (u1) edge (v1);
    \path                     (u2) edge (v2);
  \end{tikzpicture}
  \caption{Movement of strong node with restoration.}
  \label{fig:movement-restoration}
  }%
  \hspace{\fill}%
  \parbox[b]{0.45\textwidth}{%
  \centering
  \begin{tikzpicture}[style=dynnets style]
    \tikzstyle{every node}=[minimum size=0.7cm]
    
    \node[circle,draw,style=strong node]    (u3-1) at (-0.75,0)     {$u_3$};
    \node[circle,draw,style=strong node]    (u2-1) [left of=u3-1,node distance=1.0cm]   {$u_2$};
    \node[circle,draw,style=strong node]    (u1-1) [left of=u2-1,node distance=1.0cm]   {$u_1$};
    
    \draw[->] (0,0) -- (0.5,0);
    
    \node[circle,draw,style=strong node]    (u1-2) at (1.25,0)      {$u_1$};
    \node[circle,draw,style=strong node]    (u2-2) [right of=u1-2,node distance=1.0cm]  {$u_2$};
    \node[circle,draw,style=strong node]    (u3-2) [right of=u2-2,node distance=1.0cm]  {$u_3$};
    \node[circle,draw]                      (v)    [above of=u2-2,node distance=1.0cm]  {$v$};
    
    \path (u1-2)  edge  (v);
    \path (u2-2)  edge  (v);
    \path (u3-2)  edge  (v);
  \end{tikzpicture}
  \caption{Creation of a new node by a set $U=\{u_1,u_2,u_3\}$ of strong nodes.}
  \label{fig:creation}
  }%
\end{figure}%

Before stating our results, let us sketch informally but in a little more detail the definitions of Constructor's actions, which she chooses from a given set of rules.
We distinguish three different types of rules.
The first is concerned only with the ``information flow'' through the network evoked by the users; nodes and edges as such stay fixed.
A natural way to describe this aspect is to assume a labeling of the nodes that may change over time.
For instance, the label~$a$ on node~$u$ and a blank label on the adjacent node~$v$ are modified to the blank label on~$u$ and the label~$a$ on~$v$, corresponding to a shift of the data~$a$ from $u$ to~$v$.
Only the labels of adjacent non-deleted nodes can change; for the same reason as in communication networks only neighboring active clients are able to send or receive messages.
A \emph{relabeling rule} describes in which way Constructor may change labels of adjacent nodes.
The other rules entail a changes to the network structure.
Constructor can restore nodes, which could have existed before, or create completely new nodes.
These steps involve the concept of ``strongness'' of a node: a strong node cannot be deleted, and it is the prerequisite for performing the restoration and creation of nodes.
One can view strong nodes as maintenance resources of suppliers, which are located on some places in the network.
The strongness property may be propagated through edges to existing nodes; being at some node~$u$ it can also be used to restore a deleted node~$v$ by moving to it if there was an edge $(u,v)$ in the network before the deletion of~$v$ (see \Fig\ref{fig:movement-restoration}).
For the creation of a node we can pick some set~$S$ of strong nodes, create a new node~$v$ (that is or is not strong) and connect it by an edge with each node in~$S$ (see \Fig\ref{fig:creation}).
This corresponds to the assumption that creation is ``more expensive'' than restoration.
In our framework, both actions are either feasible in general or subject to constraints given by the labels of the involved nodes; we will collect these constraints in \emph{movement} and \emph{creation rules}.

Our results clarify the solvability of these \emph{connectivity games} in several versions, complementing a result of~\cite{RadTho07} that the reachability problem (to reach a connected network by Constructor) is undecidable in the cases where the creation of new nodes is allowed.
Our first results say that for safety games solvability (for Constructor) is undecidable in the general case but --~maybe surprisingly~-- decidable when either creation of new non-strong nodes is disallowed or movement of strong nodes is disallowed.
The former problem is \PSPACE-complete, the latter is solvable in \EXPTIME{} (where the input size is given by the size of the initial network and the size of the rule set of Constructor).
Also in those restrictions where creation of nodes or relabeling is completely disallowed we sharpen the results of~\cite{RadTho07} (where \PSPACE-hardness was shown) in obtaining \PSPACE-completeness.

For the last mentioned restriction (no creation of nodes, no relabelings), we finally obtain partial results by providing lower and upper bounds; the proofs illustrate again the difference between safety and reachability.
We show that under the first restriction solvability by Constructor is \PSPACE-hard (while the easily obtained upper bound is \EXPTIME{}), and for the second restriction (no relabelings) this solvability is in \PSPACE{} and \NP-hard.
An overview of the results is given in \Table\ref{tabular:connectivity-games-results}.%
\begin{table}
  \centering
  \begin{tabular}{lll} \toprule
    \multirow{2}{*}{\parbox{8em}{\textbf{game variant}\\ \hspace*{0.5em} allowed rules}}
                                            & \multicolumn{2}{c}{objective}                       \\ \cmidrule(l){2-3}
                                            & reachability                  & safety              \\ \midrule
    \textbf{expanding}                      &                               &                     \\
    \enspace{} all                          & undecidable                   & undecidable         \\
    \enspace{} s-create, w-create, relabel  & undecidable                   & in~\EXPTIME         \\
    \enspace{} s-create, move, relabel      & undecidable                   & \PSPACE-complete    \\
    \enspace{} w-create, move               & undecidable                   & open                \\
    \textbf{non-expanding}                  & \PSPACE-hard / in~\EXPTIME{}  & \PSPACE-complete    \\
    \textbf{unlabeled}                      & \NP-hard / in~\PSPACE{}       & \PSPACE-complete    \\ \bottomrule
  \end{tabular}
  \caption{Summary of our results; for expanding games, we distinguish whether Constructor is allowed to create strong nodes (\emph{s-create}), create weak nodes (\emph{w-create}), \emph{move} strong nodes, or \emph{relabel} nodes.}\label{tabular:connectivity-games-results}
\end{table}

The paper is structured as follows.
In the sequel of this introduction we discuss related but technically different approaches to network games.
Section~\ref{sect:dynamic-connectivity-game-model} introduces the model in detail.
Section~\ref{sect:connectivity-safety-games} offers the results on safety games, whereas in \Sect\ref{sect:connectivity-reachability-games} the solvability of reachability games is studied.
We finish by listing some selected perspectives from a very rich landscape of problems that remain to be treated in this area combining practical issues with theoretical research.

\paragraph{Related Work.}

Our game model was introduced in~\cite{RadTho07}; we sharpen and extend these preliminary decidability and complexity results.
The game-theoretic approach is inspired by \emph{sabotage games}, which van Benthem suggested in~\cite{Ben05}.
There, a reachability problem over graphs is considered, where a ``Runner'' traverses a graph while a ``Blocker'' deletes an edge after each move.
The theory of these games and many variants have been thoroughly studied in~\cite{LR03a,Roh05,GieKurVel09,KleRadTho10-scp}.

Dynamically changing systems are also addressed by \emph{online algorithms} (see~\cite{FiatGerhard96,BorodinElY98}).
These find applications in routing and scheduling problems in wireless and dynamically changing wired networks (see~\cite{Rajaraman02,Scheideler02}).
However, the only approaches we are aware of where the adversarial also changes the network structure is due to Awerbuch et~al.~\cite{AwerbuchBBS01,AwerbuchBS03}; there a routing objective is faced with an adversary that injects packets and also decides which connections are available.
These studies aim at a competitive analysis of the ``communication throughput'': the number of delivered packets of an online algorithm is compared to an optimal offline algorithm.

Another view on online algorithms are \emph{dynamic algorithms} (see~\cite{EppsteinGalIta99,FeigenbaumKannan00}).
A \emph{fully dynamic algorithm} refers to a dynamic graph in which edges are inserted and deleted;
the focus of investigation is the computational complexity of static graph properties with respect to a given sequence of update steps (see~\cite{HolmLT01,RodittyZwick04}).
The same idea leads to a \emph{dynamic complexity theory}, which deals with the complexity of computing and maintaining an auxiliary structure;
this structure entails the solution of a decision problem for a dynamically changing instance (see~\cite{WeberSchwentick07}).

Studies on a game-theoretic model for routing under adversarial condition have been started in~\cite{GroRadTho10}.
Instead of a competitive analysis of a given online algorithm, the aim is to check whether a given dynamic scenario has a solution in form of a routing scheme (and to synthesize a routing scheme if it exists).
This model is also inspired by the sabotage game model, but complementary to the present work.
The adversary deactivates edges and injects packets in the network, and a solution of the game requires that all packets must be delivered or that the overall number of packets in the network is bounded.

Another interesting approach arises from the studies of dynamic versions of the \emph{Dynamic Logic of Permission} ($\text{DLP}$), which is in turn an extension of the \emph{Propositional Dynamic Logic} ($\text{PDL}$).
In $\text{DLP}$, ``computations'' in a Kripke structure from one state to another are considered which are subject to ``permissions''~\cite{PucWei04}.
The logic $\text{DLP}_\text{dyn}^{+}$ (see \cite{Dem05,Goe06}) extends $\text{DLP}$ with formulas which allow updates of the permission set and thus can be seen as a dynamically changing Kripke structure.
Nevertheless, all the dynamics have to be specified in the formula; an adversarial agent is not considered.

The idea of changing networks is of course studied in considerable depth in the theory of graph grammars, graph rewriting, and graph transformations (see \cite{Corradini01,Roz97-graph-book}).
While there the class of generable graphs (networks) is the focus of study, we deal here with the more refined view when considering the evolvement of a two-player game and the properties of graphs occurring in them.
In the (one-player) framework of model checking, we mention the work~\cite{GadducciHeckelKoch98}, where \emph{graph-interpreted temporal logic} is introduced as a rule-based specification.
A technique is developed to map a ``graph transition systems'', which nodes are graphs, to a finite Kripke structure, so that classical LTL model checking can be applied.

\section{Dynamic Networks via Games}\label{sect:dynamic-connectivity-game-model}

We present \emph{networks} in the form $G = (V, E, A, S, (P_a)_{a\in\Sigma})$ with
\begin{itemize}
  \item a finite set $V$ of vertices (also called nodes),
  \item an undirected edge relation $E$,
  \item a set $A \subseteq V$ of \emph{active nodes},
  \item a set $S \subseteq A$ of \emph{strong nodes},
  \item a partition of $V$ into sets $P_a$ for some label alphabet $\Sigma$.
  A node belongs to $P_a$ if it carries the label~$a$.
\end{itemize}
We say that a node is \emph{deactivated} or \emph{deleted} if it is not active.
A \emph{weak node} is an active node which is not strong.
A network is connected if the graph that is induced by the active vertices is connected, i.e., for any two active vertices $u,v$ there exists a path from~$u$ to~$v$ which only consists of active nodes.

The dynamics of a network arises by the possible moves of two players, \emph{Destructor} and \emph{Constructor}, which are changing the respective current network.
A \emph{dynamic network game} will be presented as a pair $\cG = (G,R)$ consisting of an \emph{initial network}~$G$ as above and a finite set~$R$ of rules for Constructor.
The two players play turn by turn in alternation; Destructor starts.
Both players are allowed to skip as well.

Let us describe the rules that define the players' possible actions.
When it is Destructor's turn, he can perform a \emph{deletion step} by deleting some weak node $v \in A \setminus S$; the set $A$ is changed to $A \setminus \{v\}$.
When it is Constructor's turn, she can choose a rule from her rule set~$R$ that is applicable on the current network.
The rules in~$R$ can be of three different types.
\begin{description}
  \item[Relabeling rule:]
  A rule $\lr a,b \relabel c,d \rr$ allows Constructor to change the labels $a$ and~$b$ of two active adjacent nodes of~$A$ into $c$ and~$d$, respectively.
  Formally, for two vertices $u \in P_a$ and $v \in P_b$ with $(u,v) \in E$ the sets $P_a$, $P_b$, $P_c$, and $P_d$ are updated to $P_a \setminus \{u\}$, $P_b \setminus \{v\}$, $P_c \cup \{u\}$, and $P_d \cup \{v\}$.
  
  For relabeling rules we will also consider rules with multiple relabelings in one turn.
  This corresponds to our intuition that there can be a lot of information flow in the network at the same time.
  For example, for two relabeling steps in one turn we use the notation $\lr a,b \relabel c,d \;;\; e,f \relabel g,h \rr$.
  The relabelings are applied one after the other, but in the same turn.
  \item[Movement rule:]
  A rule $\lr a \move b \rr$ allows Constructor to shift the ``strongness'' from a strong node that carries the label~$a$ to an adjacent node that is labeled with~$b$ and must not be strong.
  Formally, for two vertices $u \in P_a$ and $v \in P_b$ with $u \in S$, $v \not\in S$, and $(u,v) \in E$, the set $S$ is updated to $(S \setminus \{u\}) \cup \{v\}$ and $A$ is updated to $A \cup \{v\}$.
  The case $v \in A$ means to simply shift strongness to~$v$; the case $v \in V \setminus A$ means \emph{restoration} of~$v$.
  The terms ``moving a strong node'' and ``shifting its strongness'' are used interchangeably through the paper.
  \item[Creation rule:]
  These rules enable Constructor to create a completely new node, which is not in~$V$.
  A rule $\lr a_1,\ldots,a_n \create[(c)] a'_1,\ldots,a'_n \rr$ allows Constructor to choose any set $U = \{u_1,\ldots,u_n\} \subseteq S$ of $n$ different strong nodes such that the label of $u_i$ is $a_i$ (for all $i \in \{1,\ldots,n\}$).
  Then, Constructor creates a new active node $w$, labels it with~$c$, and connects it to every node in~$U$.
  Formally, the sets $V$ and~$A$ are updated to $V \dotcup \{w\}$ and $A \dotcup \{w\}$, respectively; also $E$ is updated by adding edges between $w$ and each node of~$U$.
  Also the labels of the nodes in~$U$ may change after creation; the label of $u_i$ is changed to $a'_i$ (for all $i \in \{1,\ldots,n\}$).
  For the \emph{creation of a strong node} we use the notation $\lr a_1,\ldots,a_n \createstrong[(c)] a'_1,\ldots,a'_n \rr$.
  In this case also $S$ is updated to $S \dotcup \{w\}$.
\end{description}
We consider some variants where Constructor's moves are restricted.
A game $(G,R)$ is called \emph{non-expanding} if $R$ does not contain any creation rule.
In \emph{unlabeled} games nodes cannot be distinguished by theirs labels;
formally, we assume that all vertices are labeled with a blank symbol~$\blank$ and the movement rule $\lr \blank \move \blank \rr$ is the only available rule.

A \emph{play} of a game~$\cG$ is an infinite sequence $\pi = G_1 G_2 \cdots$ where $G_1$ is the initial network and each step from $G_i$ to $G_{i+1}$ results from the moves of Destructor (if $i$ is odd) and Constructor (if $i$ is even), respectively.
So, plays are infinite in general, but may be considered finite in the cases where neither of the players can move, or a given objective (winning condition) is satisfied.
In this paper we consider \emph{dynamic network connectivity games}, where Constructor's objective concerns the connectivity of the network (more precisely, of the active nodes).
We distinguish between connectivity games with \emph{reachability objective} and \emph{safety objective}.
In the former the initial network is disconnected, and Constructor's objective is to reach a connected network;
conversely, in the safety game the initial network is connected, and Constructor has to guarantee that the network always stays connected.
If Constructor achieves the given objective in a play~$\pi$, she \emph{wins~$\pi$} with respect to this objective.

A \emph{strategy for Destructor} is a function (usually denoted by~$\sigma$) that maps each play prefix $G_1 G_2 \cdots G_k$ with an odd~$k$ to a network $G_{k+1}$ that arises from $G_k$ by a node deletion.
Analogously, a \emph{strategy for Constructor} is such a function (denoted by~$\tau$) where $k$ is even and $G_{k+1}$ arises from $G_k$ by applying one of the rules from~$R$.
A strategy is called \emph{positional} (or \emph{memoryless}) if it only depends on the current network, i.e., it is a function that maps the current network $G_k$ to~$G_{k+1}$ as above.
If Constructor has a strategy~$\tau$ to win every play, where she moves according to~$\tau$, with respect to the reachability (safety) objective, \emph{Constructor wins the reachability (safety) game}; Destructor wins otherwise.
This leads us to the following decision problems.
\begin{itemize}
  \item \emph{Dynamic reachability problem:} Given a dynamic network game~$\cG$, does Constructor have a strategy to win the reachability game~$\cG$ (i.e., eventually a connected network is reached)?
  \item \emph{Dynamic safety problem:} Given a dynamic network game~$\cG$, does Constructor have a strategy to win the safety game~$\cG$ (i.e., the network always stays connected)?
\end{itemize}

In this paper we only consider reachability and safety objectives.
For these winning conditions it is well known that one can restrict winning strategies of both players to positional strategies~\cite{Thomas95,GTW02}, i.e., if Constructor (Destructor) has a strategy to win a given game~$\cG$, she (he) also has a positional strategy to win~$\cG$.
Therefore, we will always assume positional strategies in this paper.

\begin{example}
We consider a dynamic network connectivity game $\cG = (G,R)$ with labels $\bot, \blank \in \Sigma$.
The initial network $G = (V,E,A,S,(P_a)_{a\in\Sigma})$ with the set $S = \{s_1,s_2,u_1,w_1\}$ of strong nodes is depicted in \Fig\ref{fig:example}.
We consider the safety game where the only rule in~$R$ is $\lr \blank \move \blank \rr$.
It means that the strong nodes $s_1$ and~$s_2$ are not able to move, because their labels do not match the movement rule.
Constructor has to guarantee the connectivity of the network. 
As a scenario for this game one could imagine two clients $s_1, s_2$ communicating over a network with unreliable intermediate nodes but two mobile maintenance resources (initially located on $u_1$ and~$v_1$).%
\begin{figure}%
  \centering
  \begin{tikzpicture}[style=dynnets style,label distance=-0.33ex]
    \tikzstyle{every node}=[minimum size=0.7cm]
    
    \node[circle,draw,label=below:$u_1$,style=strong node]  (u1)                        {$\blank$};
    \node[circle,draw,label=above:$u_2$]                    (u2)  [above of=u1]         {$\blank$};
    \node[circle,draw,label=above:$u_3$]                    (u3)  [left of=u2]          {$\blank$};
    \node[circle,draw,label=above:$v_1$]                    (v1)  [right of=u2]         {$\blank$};
    \node[circle,draw,label=below:$v_2$]                    (v2)  [below of=v1]         {$\blank$};
    \node[circle,draw,label=above:$w_1$,style=strong node]  (w1)  [right of=v1]         {$\blank$};
    \node[circle,draw,label=below:$w_2$]                    (w2)  [below of=w1]         {$\blank$};
    \node[circle,draw,label=below:$w_3$]                    (w3)  [right of=w2]         {$\blank$};
    \node[circle,draw,label=below:$s_1$,style=strong node]  (s1)  [left of=u1]          {$\bot$};
    \node[circle,draw,label=above:$s_2$,style=strong node]  (s2)  [right of=w1]         {$\bot$};
    
    \path (u1) edge (u2) edge (u3)    (u2) edge (u3)
          (w1) edge (w2) edge (w3)    (w2) edge (w3)
          (v1) edge (u2) edge (w1)
          (v2) edge (u1) edge (w2)
          (s1) edge (u1) edge (u3)
          (s2) edge (w1) edge (w3);
  \end{tikzpicture}
  \caption{An example initial network (bold nodes are strong).}
  \label{fig:example}
\end{figure}
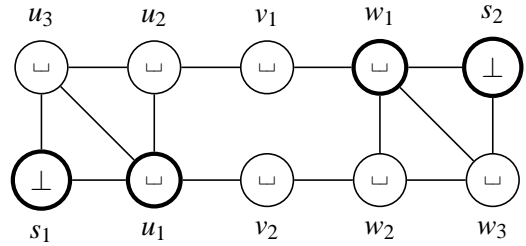

By taking a closer look at this example we see that Destructor has a winning strategy.
He deletes $w_3$ in his first move; then we distinguish between two cases:
if Constructor restores $w_3$, Destructor deletes $v_1$ in his next move and finally $u_1$ or~$v_2$;
if Constructor does not move the upper movable strong node to~$w_3$, this node has to remain at~$w_1$; otherwise Constructor loses by deletion of~$w_1$.
In the second case it is easy to see that Destructor wins by suitable deletions of nodes in $\{u_1,u_2,v_1,v_2\}$.

Now we consider the same game, but additionally with the creation rule $\lr \blank,\blank \create[(\blank)] \blank,\blank \rr$.
We claim that now Constructor has a winning strategy.
If Destructor deletes the node $v_1$ or~$v_2$ Constructor creates a new vertex~$v_3$ with the creation rule which establishes a new connection between the two strong nodes $u_1$ and~$w_1$.
If Destructor deletes the new vertex~$v_3$ Constructor creates a new vertex again, and so on.
Note that in this way the number of vertices in the set~$V$ can increase to an unbounded number.
\end{example}

\section{Results for Safety Connectivity Games}\label{sect:connectivity-safety-games}

\paragraph{The general case.}
In this section we analyze the dynamic safety problem, for which we show in our first result that it is undecidable in general.
It is indeed remarkable that we have to assume the presence of weak creation, movement, and relabeling rules to show this.
Later we will see that the dynamic safety problem becomes decidable if weak creation or movement rules are absent.
\begin{theorem}\label{thm:connectivity-safety-games-undecidable}
  The dynamic safety problem is undecidable, even if Constructor can only apply weak creation, movement, and relabeling rules.
\end{theorem}
\begin{proof}
  We use a reduction from the halting problem for Turing machines.
  These are w.l.o.g.\ presented in the format $M = (Q,\Gamma,\delta,q_0,q_\text{stop})$ with a state set~$Q$, a tape alphabet~$\Gamma$ (including a blank symbol~$\blank$),
  a transition function $\delta: Q \setminus \{q_\text{stop}\} \times\Gamma \rightarrow Q\times\Gamma\times\{L,R\}$,
  an initial state~$q_0$, and a stop state~$q_\text{stop}$.

  For a Turing machine~$M$ we construct a game $\cG = (G,R)$ such that $M$ halts when started on the empty tape iff Constructor is not able to keep the network always connected by applying the rules of~$R$, i.e., Destructor wins the safety game~$\cG$.
  The idea is to consider a configuration of~$M$ as a connected network where Constructor creates additional vertices during the simulation of a valid computation of~$M$.
  If $M$ stops, she cannot create vertices anymore, and Destructor is able to disconnect the network.
  We label the nodes that correspond to a configuration of~$M$ with triples of the from $\Gamma \times (\hat{Q} \cup \{L,R\}) \times \{|,]\}$ with $\hat{Q} \coloneqq Q \times \{0,1,\triangleleft,\triangleright\}$.
  The first component holds the content of its represented cell of the tape.
  The second component is labeled with $L$ or~$R$ if the represented cell is on the left-hand side of the head or on the right-hand side of the head, respectively;
  the second component is labeled with $q \in Q$ and some auxiliary symbol if $M$ is in state~$q$ and the head is on the cell represented by this node.
  The third element is either an end ($\,]\,$) or an inner marker ($\,|\,$) depending on whether the node is the currently the right-most represented cell of the tape or not.
  Additionally, the label alphabet contains the symbols $\top$, $\bot$, $+$, and~$!$.
  The labels $\top$, $\bot$ are used for the two additional strong nodes that Constructor has to keep connected; the $\bot$-labeled node is always connected to every cell node while the $\top$-labeled node is only connected to the $\bot$-labeled node via some weak nodes that are labeled with~$+$.
  The exclamation mark ($!$) is used as a label that Destructor has to prevent to occur; if Constructor manages to relabel a strong node to a~$!$-labeled node, she has a winning strategy regardless of the behavior of~$M$.
  
  Constructor has to create a $+$-labeled weak node in every turn where she simulates a transition of~$M$.
  Since we want Constructor to simulate only valid transitions, we ensure that an according creation rule can only be applied to the cell node that holds the current state of~$M$ and an adjacent cell node.
  For this reason only two cell nodes are strong at any time.
  Constructor is able to shift these nodes depending on whether Constructor wants to simulate a left or a right transition of~$M$.
  We ensure that Constructor shifts the nodes at most once between simulating two transitions; otherwise she would be able to shift them forever instead of simulating~$M$.
  For this reason the cell node representing the head has auxiliary symbols in $\{0,1,\triangleleft,\triangleright\}$:
  the symbol~$0$ means that Constructor can choose either to shift the strong nodes or to simulate a transition.
  If this symbol is~$1$, she has already shifted the strong nodes and now must simulate a transition.
  The symbols $\triangleleft$ and~$\triangleright$ are used as intermediate labels when Constructor shifts the strong nodes to the left or to the right, respectively.
  The initial network, which corresponds to the initial configuration of~$M$ on an empty working tape, is depicted in \Fig\ref{fig:safety-undicadable-initial-network}.%
  \begin{figure}%
    \centering
    \begin{tikzpicture}[style=dynnets style]
      \tikzstyle{every node}=[minimum size=0.7cm]
      
      \draw (0,0)       node[circle,draw,style=strong node] (top)  {$\top$};
      \draw (0,-1)      node[circle,draw]                   (aux1) {+};
      \draw (-1.2,-1)   node[circle,draw]                   (aux2) {+};
      \draw (+1.2,-1)   node[circle,draw]                   (aux3) {+};
      \draw (0,-2)      node[circle,draw,style=strong node] (bot)  {$\bot$};
      \draw (-1.5,-3.2) node[rectangle, draw, rounded corners, style=strong node, minimum width=2.2cm] (u1) {$(\blank,(q_0,0),|)$};
      \draw (+1.5,-3.2) node[rectangle, draw, rounded corners, style=strong node, minimum width=2.2cm] (u2) {$(\blank,R,])$};
      
      \path (top) edge (aux1);
      \path (top) edge (aux2);
      \path (top) edge (aux3);
      \path (bot) edge (aux1);
      \path (bot) edge (aux2);
      \path (bot) edge (aux3);
      \path (bot) edge (u1);
      \path (bot) edge (u2);
      \path (u1)  edge (u2);
    \end{tikzpicture}
    \caption{A game network representing an initial configuration of~$M$.}
    \label{fig:safety-undicadable-initial-network}
  \end{figure}
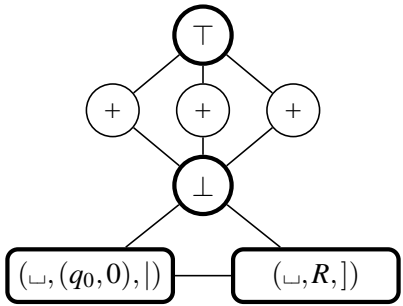
  
  In the following we describe the rule set~$R$.
  As mentioned before, the rule $\lr !,\top,\bot \create[(+)] !,\top,\bot \rr$ allows Constructor to ensure the connectivity of the network if a strong node obtains the $!$-label.
  
  To allow Constructor to shift the two strong cell nodes to the right, we add the following rules for all $q \in Q$, $a,b \in \Gamma$, and $* \in \{|,]\}$:
  1.~$\lr (a,(q,0),|), \top, \bot \create[(+)] (a,(q,\triangleright),|), \top, \bot \rr$,
  2.~$\lr (a,(q,\triangleright),|) \move (b,R,*) \rr$,
  3.~$\lr (a,L,|) \move (b,(q,\triangleright),|) \rr$,
  and 4.~$\lr (a,(q,\triangleright),|,0), \top, \bot \create[(+)]$ $(a,(q,1),|), \top, \bot \rr$.
  The rules for shifting the two strong nodes to the left are built analogously.
  After the application of the second and the third rule, we want to force Destructor to deactivate the weak cell node (instead of a $+$-labeled node).
  For this reason we add the relabeling rule
  $\lr (a,L,|),(b,(q,z),|) \relabel \protect{!,!} \;;\; \protect{!,(c,R,*)} \relabel \protect{!,!} \rr$ for every $a,b,c \in \Gamma$, $z \in \{\triangleleft,\triangleright\}$, and $* \in \{|,]\}$.
  An application of this rule is possible iff a series of three cell nodes is active; it leads to an $!$-labeled strong node and hence to a network where Constructor wins.
  
  A transition of~$M$ is simulated by changing the labels of the two strong cell nodes.
  One of the cell nodes has to carry, besides the state of~$M$, the auxiliary symbol $0$ or~$1$; in this case it is guaranteed that the two strong cell nodes are adjacent.
  Due to the introduced move rules, we can assume that these strong nodes are already at their desired position.
  Then, it is easy to supply a set of creation rules that mimics the transitions of~$M$.
  Formally, for each tuple $(q,a,p,b,X)$ with $\delta(q,a)=(p,b,X)$ and for every $c \in \Gamma$, $z \in \{0,1\}$, and $* \in \{|,]\}$ we add the rule
  $\lr (c,R,|),(a,(q,z),*),\top, \bot \create[(+)] (c,(p,0),|),(b,R,*),\top, \bot \rr$ if $X=L$, and
  $\lr (a,(q,z),|),(c,R,*),\top, \bot \create[(+)] (b,L,|),(c,(p,0),*),\top, \bot \rr$ if $X=R$.
  
  Finally, rules are needed to extend the network in the case that more space on the tape is needed.
  New cell nodes are allocated next to the end marker, which represents the rightmost used cell of the tape.
  For this allocation we add the rule
  $\lr \bot,(a,(q,0),]) \create[((\blank,R,{]}))] \bot,(a,(q,0),|) \rr$ for every $a \in \Gamma$, and $q \in Q$.
  Destructor will deactivate the created node with the label $(\blank,R,{]})$ immediately to prevent Constructor from relabeling a strong tape node to a $!$-labeled node.
  
  To show the correctness of the construction, assume that $M$ never stops.
  In this case Constructor can guarantee that there is at least one active $+$-labeled node, which connects the nodes labeled $\top$ and~$\bot$.
  One of the three $+$-labeled nodes in the initial network is deleted since Destructor starts.
  Destructor may delete another of these nodes if he misbehaves after some tape extension or a strong node shift and hence allows Constructor to obtain a strong $!$-labeled node.
  In each case Constructor wins.
  Conversely, if $M$ stops, Constructor cannot apply any rule for simulating a transition from some point onwards.
  The construction ensures that Constructor can shift the strong cell nodes or create a new cell node at most once after simulating a transition.
  Hence, she can only skip from some point onwards, and Destructor wins by deleting all $+$-labeled nodes.
\end{proof}

\paragraph{Decidable Subcases.}
Now, we analyze safety games under some restrictions to the given rule set.
If we prohibit weak creation rules, solving safety games is \PSPACE{}-complete.
The \PSPACE-hardness also holds in the more restricted unlabeled case (see \Theorem\ref{thm:unlabeled-connectivity-safety-games-pspace-hard}).
Here, we show the inclusion in \PSPACE{}.

We call a strategy of Destructor \emph{strict} if he deletes a vertex in every turn (i.e., he does not skip) whenever there is still a weak node left for deletion.
We can assume that Destructor always plays a strict strategy in a safety game:
if Destructor skips, so Constructor can skip as well leading the play to the same network (which is still connected).
\begin{remark}\label{remark:connectivity-safety-games-strict-strategies}
  If Destructor wins a safety game~$\cG$, he also has a strict strategy to win~$\cG$.
\end{remark}

For a play $\pi = G_1 G_2 \ldots$ we define the \emph{level} of a network~$G_i$ as the number of weak nodes in~$G_i$ if Destructor acts next (i.e., $i$ is odd) and as the number of weak nodes in~$G_i$ minus~$1$ if Constructor moves next (i.e., $i$ is even).
Clearly, if Destructor plays according to a strict strategy, the level is monotonically decreasing as long as the level has not reached~$0$ (or Destructor has won).

\begin{lemma}\label{lemma:connectivity-safety-games-bound-for-strongness-moves}
  Consider a safety game~$\cG$ without weak creation rules.
  If Destructor wins~$\cG$, he also wins~$\cG$ with a strict strategy such that, for each~$\ell$, Constructor is able to shift each strongness at most $n_\ell \cdot d_\ell$ times in networks of level~$\ell$ (before a disconnected network is reached), where $n_\ell$ ($d_\ell$) is the number of nodes (deactivated nodes) of the first occurring network of level~$\ell$.
\end{lemma}
\begin{proof}
  Towards a contradiction, assume that Destructor has a strict winning strategy~$\sigma$, but Constructor has a strategy~$\tau$ where, for some~$\ell$, she is able to shift a strongness more than $n_\ell \cdot d_\ell$ times in networks of level~$\ell$ before Destructor wins.
  Consider a play~$\pi$ where Destructor and Constructor play according to $\sigma$ and~$\tau$, respectively.
  So, there exists some $\ell$ such that Destructor shifts a strongness at least $n_\ell \cdot d_\ell + 1$ times in networks of level~$\ell$.
  Let $G_i$ be the first network of level~$\ell$ in~$\pi$, and let $G_k$ be the last network of level~$\ell$ in~$\pi$, where either Destructor has already won (i.e., $G_k$ is disconnected) or Constructor's move decreases the level.
  Since weak creation rules, which preserves the level, are forbidden and Destructor's strategy~$\sigma$ is strict, we can assume that Constructor applies only movement rules in the play infix $G_i \cdots G_k$ and hence the set of nodes and their labels are preserved in this play infix; applying a relabeling or a strong creation rule would immediately decrease the level to~$\ell-1$.
  
  In the play infix $G_i \cdots G_k$ each strongness is shifted along a certain path of nodes, each of which must have been deactivated before Constructor shifts the strongness to it; otherwise the level would decrease to~$\ell-1$ immediately.
  Among these deactivated vertices we distinguish, for each network in $G_i \cdots G_k$, between the nodes that have already been deactivated since~$G_i$ and the nodes that have been deleted by Destructor in some network of level~$\ell$ at least once.
  As the network~$G_i$ consists of $d_\ell$ deactivated nodes, in the play infix $G_i \cdots G_k$ Constructor shifts a strongness at most $d_\ell$ times to a node that has not been deleted by Destructor in some network of level~$\ell$ before.
  Since there is a strongness that Constructor shifts at least $n_\ell \cdot d_\ell + 1$ times in networks of level~$\ell$, there is a play infix $G_{j_1} \cdots G_{j_2}$ of~$\pi$ with $i \leq j_1 < j_2 \leq k$ where a strongness is shifted in a loop such that the node where this strongness is shifted to has been deleted by Destructor before in some network of level~$\ell$.
  Assume that this loop consists of $m$ nodes.
  
  Since Constructor restores these $m$ nodes, none of these $m$ nodes stay deactivated until Destructor wins or the level decreases.
  It remains to be shown that Destructor does not have to delete all of these $m$ nodes in order to prevent Constructor from applying a certain rule.
  By definition the $m$ deleted nodes are restored by the same strongness; none of the other strong nodes has to be moved in order to restore them.
  The vertices, edges, and labels of the network stay unchanged during the loop.
  So, Constructor's possibilities for node creation and movement are not constricted.
  In the case we assume that Destructor must delete all of the $m$ nodes to prevent Constructor from applying a relabeling rule, we obtain a winning strategy for Constructor since she would be able to move the strong node in the loop again and again, which would take her as many turns as Destructor needs for the node deletions (in this case Destructor would not be able to perform any other node deletion).
  
  Therefore, at least one of these $m$ node deletions is needless for Destructor; we can eliminate it from Destructor's strategy without harming his strict winning strategy.
  (For the elimination step, we let Destructor successively delete the next weak node that he would delete by playing his strategy~$\sigma$.)
  Repeating this elimination for every case where Constructor can shift some strongness in networks of the same level in a loop, we can optimize Destructor's strict winning strategy to one where he additionally prevents for all~$\ell$ that any strongness is shifted more than $n_\ell \cdot d_\ell$ times in networks of level~$\ell$.
\end{proof}

So, for safety games where Destructor wins we obtained an upper bound of the length of the path along that a strongness is shifted within the same level.
From this we can derive an upper bound for the number of node deletions that Destructor needs if he has a winning strategy.
\begin{lemma}\label{lemma:no-weak-create-connectivity-safety-games-bound}
  Consider a safety game~$\cG$ without weak creation rules.
  Let $|V|$ ($|S|$) be the number of active nodes (strong nodes) of the initial network.
  If Destructor wins~$\cG$, he also has a strict strategy to win~$\cG$ with at most $|S| \cdot (2|V|-|S|)^3$ node deletions.
\end{lemma}
\begin{proof}
  Assume that Destructor wins the safety game~$\cG$.
  The previous lemma states that Destructor also wins with a strict strategy where, for each~$\ell$, Constructor can shift each strongness at most $n_\ell \cdot d_\ell$ times in networks of level~$\ell$.
  Since the number of strong nodes is fixed, Destructor wins with a strict strategy where, for each~$\ell$, he acts at most $|S| \cdot n_\ell \cdot d_\ell = |S| \cdot n_\ell \cdot (n_\ell-|S|-\ell)$ times in networks of level~$\ell$.
  For strict strategies the level is monotonically decreasing (as long as it has not reached~$0$).
  The level decreases at most $|V|-|S|$ times; so, $n_\ell \leq |V| + (|V|-|S|) = 2|V|-|S|$ for every~$\ell$.
  Hence, Destructor wins with a strict strategy deleting at most
  $\sum_{\ell=0}^{|V|-|S|} |S| \cdot n_\ell \cdot (n_\ell-|S|-\ell) \leq |S| \cdot (2|V|-|S|)^3$
  nodes.
\end{proof}

To show that the dynamic safety problem is in \PSPACE{} (if weak creation rules are forbidden) it suffices to build up the game tree, which we truncate after $|S| \cdot (2|V|-|S|)^3$ moves of Destructor.
We construct the game tree on-the-fly in a depth-first manner, so that we only have to store a path from the root to the current node, which length is polynomial in the size of~$\cG$.
\begin{theorem}\label{thm:no-weak-create-safety-games-decidable}
  For games where Constructor does not have any weak creation rule, the dynamic safety problem is in \PSPACE.
\end{theorem}

Another decidable subclass of the dynamic safety problem is the case where Constructor cannot move any strong node.
Since Constructor is not able to restore any deleted node, we can ignore the deleted nodes in each network.
Hence, we only have to explore are finite state space.
\begin{theorem}\label{thm:no-move-safety-games-decidable}
  For games where Constructor does not have any movement rule, the dynamic safety problem is in \EXPTIME.
\end{theorem}
\begin{proof}[Proof sketch]
  Consider a safety game~$\cG$ without movement rules.
  Due to \Remark\ref{remark:connectivity-safety-games-strict-strategies} we can assume w.l.o.g.\ that Constructor plays according to a strict strategy.
  Then, the level decreases at most $|V|-|S|$ times before Destructor wins or all nodes in the network are strong,
  where $|V|$ ($|S|$) is the number of active nodes (strong nodes) in the initial network.
  We transform $\cG$ into an \emph{infinite game}~\cite{GTW02} on a game graph~$G'$, where each vertex corresponds to a network of~$\cG$ and the information which player acts next.
  When we ignore deleted nodes, the number of networks of the same level is at most exponential in~$\cG$.
  Since the number of different levels is linear in~$\cG$, the size of~$G'$ is at most exponential in~$\cG$ and can be computed in exponential time.
  The dynamic safety problem for~$\cG$ is equivalent to the problem of determining the winner in the safety game on~$G'$, which is decidable in linear time~\cite{Thomas95,GTW02}.
\end{proof}

\paragraph{Non-Expanding and Unlabeled Games.}
We already showed in \Theorem\ref{thm:no-weak-create-safety-games-decidable} that we can solve the dynamic safety problem in \PSPACE{} if weak creation rules are forbidden.
This lower bound cannot be improved in the more restricted cases of non-expanding and unlabeled games since it has already been shown in~\cite{RadTho07} that the dynamic safety problem is \PSPACE-hard for unlabeled games.
\begin{theorem}\label{thm:unlabeled-connectivity-safety-games-pspace-hard}
  For unlabeled games, the dynamic safety problem is \PSPACE-hard.
\end{theorem}
So, the dynamic safety problem is \PSPACE-complete also for unlabeled and non-expanding games.

\section{Results for Reachability Connectivity Games}\label{sect:connectivity-reachability-games}

\paragraph{The General Case.}
The dynamic reachability problem is also undecidable in general.
This has already been shown in~\cite{RadTho07} for the variant where Constructor has to reach a biconnected network (instead of a connected network).
The result also holds if we allow Constructor to use only strong creation and relabeling rules.
Moreover, in the reachability game a Turing machine can be simulated solely by Constructor who may connect a network if a stop state is reached, whereas in the safety game she has to simulate transitions in order to compensate Destructor's node deletions.
As a consequence the undecidability of the dynamic reachability problem holds even for the solitaire game version where Destructor always decides to skip.
One can easily adapt the proof of this result to the case where Constructor only has to establish a connected network.
Alternatively, we can also use the idea of the proof of \Theorem\ref{thm:connectivity-safety-games-undecidable} to relabel adjacent ``cell nodes'' with weak creation rules and guarantee with movement rules that only these two adjacent ``cell nodes'' are strong.
\begin{theorem}\label{thm:connectivity-reachability-games-undecidable}
  The dynamic reachability problem is undecidable, even if Constructor can only apply strong creation and relabeling rules or she can only apply weak creation and movement rules.
  In both cases the problem remains undecidable in the solitaire game version where Destructor never moves.
\end{theorem}

\paragraph{Non-Expanding and Unlabeled Games.}
The proofs for the undecidability of the dynamic reachability problem for expanding games rely on the availability of creation moves; if these are omitted, the state space is finite and hence the problem becomes trivially decidable.
\begin{remark}\label{remark:non-expanding-safety-in-exptime}
  For non-expanding games, the dynamic reachability problem is in \EXPTIME.
\end{remark}

Complementary to this \EXPTIME{} upper bound, we have a \PSPACE{} lower bound.
\begin{theorem}\label{thm:non-expanding-connectivity-reachability-games-pspace-hard}
  For non-expanding games, the dynamic reachability problem is \PSPACE-hard.
\end{theorem}
This result is a variant of a result in~\cite{RadTho07}, where the \PSPACE-hardness has been shown for the question of whether Constructor can reach a network in which a certain label occur.
It can be shown by a reduction from the sabotage game problem.

In the \emph{unlabeled} non-expanding case we give an \NP{} lower bound and a \PSPACE{} upper bound.
\begin{theorem}\label{thm:unlabeled-connectivity-reachability-games-np-hard}
  For unlabeled games, the dynamic reachability problem is \NP-hard.
\end{theorem}
\begin{proof}[Proof sketch]
  We use a polynomial-time reduction from the \emph{vertex cover} problem.
  The basic idea is to use a graph, say $G_\text{VC}$, as a network~$G$ where the original vertices are taken as deactivated nodes and the original edges are taken as weak intermediate nodes; moreover, $G$ consists of $k$ strong nodes, which are connected to the deactivated nodes.
  If Constructor is able to connect~$G$ by moving these strong nodes to the deactivated nodes, those (formerly deactivated) nodes form a vertex cover in $G_\text{VC}$;
  and conversely, if $G_\text{VC}$ has a vertex cover of size~$k$, Constructor wins by moving the strong nodes to this vertex cover.
\end{proof}

Now, we establish a \PSPACE{} upper bound for the unlabeled case.
The basic observation is that, if Constructor moves some strong node a certain number of times, she moves a strong node in a loop that cannot be necessary for a winning strategy.
For this purpose, we first note an upper bound on the number of moves of a strong node; we know that after $k \cdot |V|$ moves Constructor has shifted this strongness in some loop at least $k$ times starting from a certain vertex.
\begin{remark}
  If Constructor shifts some strongness $k \cdot |V|$ times, there is a vertex $v \in V$ that this strongness visits $k+1$ times, i.e., the strongness is shifted through $k$ loops that start and end at~$v$.
\end{remark}

We show that Constructor does not need to shift a strong node through more than $2 \cdot |V| - 1$ loops starting from the same vertex.
Then we can infer from the previous remark that it is sufficient for Constructor to shift each strongness at most $2 \cdot |V|^2 - 1$ times.
\begin{lemma}\label{lemma:connectivity-reach-games-bound-for-strongness-moves}
  Consider an unlabeled reachability game~$\cG$.
  If Constructor wins~$\cG$, she also wins~$\cG$ with a strategy where she shifts each strongness at most $2 \cdot |V|^2 - 1$ times.
\end{lemma}
\begin{proof}
  Towards a contradiction, we assume that Constructor has a winning strategy~$\sigma$, but that Destructor has a strategy~$\tau$ such that Constructor has to shift some strongness at least $2 \cdot |V|^2$ times before she wins.
  Consider a play~$\pi$ where Destructor and Constructor play according to $\sigma$ and~$\tau$, respectively.
  Then, the previous remark states that there is a vertex~$v \in V$ from which Constructor moves some strongness through at least $2 \cdot |V|$ loops before Constructor wins the play~$\pi$.
  
  In a reachability game where only movement rules are allowed, Destructor cannot restrict Constructor's possibilities to move.
  Hence, there are only two possible reasons for Constructor to shift the mentioned strong node in a loop that starts and ends at~$v$.
  \begin{enumerate}
    \item
    Some node of the loop, say~$u$, is restored by shifting the strong node in that loop.
    However, in this case we can assume that Constructor does not restore $u$ again while shifting the strongness in a loop that starts and ends at~$v$.
    Otherwise Constructor can omit each former loop in which she shifts the strong node only for this reason; Constructor will still win with this modified strategy.
    \item
    Destructor deletes some node $u \in V \setminus \{v\}$.
    Also in this case we can assume that Constructor does not move again this strongness in a loop that starts and ends at~$v$ while Destructor can ensure that the deletion of~$u$ is Constructor's only achievement (e.g., it may be that Destructor loses during this loop if he does not delete~$u$).
    Otherwise the deletion of~$u$ in each former loop does not let Constructor establish a connected network.
    Hence, Constructor can omit each former loop in which she shifts the strong node only for this reason; again, she will still win with this strategy.
  \end{enumerate}
  Thus, we can assume that the first case occurs at most~$|V|$ times and the second case occurs at most $|V|-1$ times if Constructor plays optimal.
  Hence, we can optimize Constructor's winning strategy~$\tau$ to a winning strategy~$\tau'$ with which she shifts each strong node through at most $2 \cdot |V| - 1$ loops that start and end at the same vertex.
  This is a contradiction to our assumption that Constructor has to shift some strongness at least $2 \cdot |V|^2$ times;
  if this was the case it would follow from previous lemma that this strongness is shifted through at least $2 \cdot |V|$ loops that start and end at the same vertex.
\end{proof}

We lift the upper bound for the number of moves of each strong node (in reachability games where Constructor wins) to the overall number of moves that Constructor needs to win.
\begin{lemma}\label{lemma:unlabeled-connectivity-reach-games-bound}
  Consider an unlabeled reachability game~$\cG$ where the network consists of $|V|$ vertices, $|S|$ of which are strong.
  If Constructor wins~$\cG$, she also has a strategy to win~$\cG$ with at most $2 \cdot |S| \cdot |V|^2 - 1$ moves.
\end{lemma}
\begin{proof}
  For connectivity games with reachability objective we can assume that Constructor never skips:
  if she skips, Destructor can skip as well leading the play to the same (disconnected) network.
  Since Constructor never skips, there exists a strongness that she shifts at least $k$ times within $|S| \cdot k$ moves.
  By \Lemma\ref{lemma:connectivity-reach-games-bound-for-strongness-moves} Constructor wins with $2 \cdot |S| \cdot |V|^2 - 1$ moves if she has a winning strategy.
\end{proof}

To show the decidability in \PSPACE{} it suffices to build up the game tree, which we truncate after $2 \cdot |S| \cdot |V|^2 - 1$ moves of Constructor (analogously to \Theorem\ref{thm:no-weak-create-safety-games-decidable}).
\begin{theorem}\label{thm:unlabeled-connectivity-reach-games-decidable}
  For unlabeled games, the dynamic reachability problem is decidable in \PSPACE.
\end{theorem}

\section{Perspectives}\label{sect:perspectives}

We have introduced dynamic network connectivity games and studied the dynamic reachability and the dynamic safety problem for them.
We showed that both problems are undecidable in general.
However, restricting the permitted rule types we pointed out decidable fragments and encountered fundamental differences in the decidability and computational complexity of the reachability the safety version of the problem.
As conclusion we mention some concrete open issues and possible refinements of the model.
\begin{enumerate}
  \item One can consider versions of the dynamic reachability and the dynamic safety problem where other subsets of the rules are allowed as considered in this paper.
  Some of these cases are trivial whereas others seem to be challenging:
  for instance the question of whether the dynamic safety problem is decidable if only relabeling rules are prohibited,
  or the question of whether the dynamic reachability problem is decidable if only creation rules are allowed.
  \item We have a gap between the upper and the lower bound for the complexity of solving non-expanding reachability games.
  We conjecture these are easier to solve in the unlabeled case as in the general non-expanding case; however, a proof is still missing.
  \item Some of our results depend on the balance of node deletion and restoration: if Constructor restores a node, Destructor can delete another one immediately.
  If one allows rules for multiple movements and relabelings in Constructor's turns, the complexity of the dynamic reachability and the dynamic safety problem increases.
  (In the non-expanding case, \EXPTIME{}-completeness can be obtained via a reduction from the halting problem of polynomial space bounded alternating Turing machines.)
  \item We focused on reachability and safety specifications for the formal analysis of networks.
  In practice one may consider a more involved \emph{recurrence (B\"uchi) condition}, where Constructor has to reach a connected network again and again, or a \emph{persistence (co-B\"uchi) condition}, where Constructor has to guarantee that the network stays connected from some point onwards~\cite{GTW02}.
  \item In the same way one may consider properties in \emph{linear temporal logic} (LTL).
  A slight generalization in the context of connectivity games are LTL-specifications over a single predicate that is true in turn~$i$ iff the current network is connected in turn~$i$.
  For non-expanding games with such an LTL-condition an \EXPTIME{} lower and a \TwoEXPTIME{} upper bound are known~\cite{Gruener2011-diploma}.
  \item It is rarely realistic to assume an omniscient adversary who deletes nodes; faults are better modeled as random events.
  This scenario has been studied in the framework of sabotage games~\cite{KleRadTho10-scp}. 
  One can study the corresponding case for dynamic network connectivity games, where Destructor is replaced by random vertex deletions~\cite{Gruener2011-diploma}.
  \item Another aspect is that yes/no questions as studied in this paper (i.e., whether a given specification is satisfied or not) have to be refined.
  From a practical point of view the formulation of an optimization problem is more useful, where we ask how many strong nodes are necessary to guarantee the network connectivity.
  For this optimization problem simple heuristics yield small (although not optimal) solutions with efficient winning strategies on various classes of networks~\cite{Gruener2011-diploma}.
\end{enumerate}

\paragraph{Acknowledgment.}
We thank a careful referee for his/her remarks on \Lemma\ref{lemma:connectivity-safety-games-bound-for-strongness-moves} and \Lemma\ref{lemma:connectivity-reach-games-bound-for-strongness-moves}; they helped to clarify the arguments.


\bibliographystyle{eptcs}
\bibliography{dynnets-refs}

\end{document}